# Shear-layer dynamics at the interface of parallel Couette flows

Manohar Teja Kalluri and  Vagesh D. Narasimhamurthy

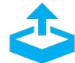

## ARTICLES YOU MAY BE INTERESTED IN

Direct numerical simulation of planar turbulent jets: Effect of a pintle orifice
Physics of Fluids **34**, 105111 (2022); https://doi.org/10.1063/5.0113460

Structure of turbulence in planar rough Couette flows
Physics of Fluids **34**, 065124 (2022); https://doi.org/10.1063/5.0092037

Multiscale modeling of different cavitating flow patterns around NACA66 hydrofoil
Physics of Fluids **34**, 103322 (2022); https://doi.org/10.1063/5.0117162

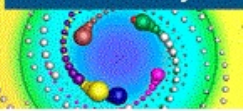

# Shear-layer dynamics at the interface of parallel Couette flows

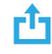

Manohar Teja Kalluri[a)] and Vagesh D. Narasimhamurthy

AFFILIATIONS

Department of Applied Mechanics, Indian Institute of Technology Madras, Chennai 600036, India

[a)]Current address: 814, Laver Building, Streatham Campus, University of Exeter, Exeter EX4 4QE, United Kingdom.
Author to whom correspondence should be addressed: manohartejakalluri25@gmail.com

ABSTRACT

This article aims to make a detailed analysis of co-flowing plane Couette flows. Particularly, the variation of flow quantities from the turbulent to non-turbulent region is studied. While the enstrophy exhibits a sharp jump, the other quantities (e.g., mean velocity, Reynolds normal stress, and kinetic energy) show a continuous variation across the interface. The budget analysis of Reynolds normal stresses reveals that the terms playing a key role in turbulence transportation vary depending on the Reynolds normal stress under study. The terms production, diffusion, and redistribution play an important role in streamwise Reynolds stress ($\overline{u'u'}$). In the spanwise Reynolds stress ($\overline{v'v'}$), the diffusion terms play a significant role. In the wall-normal Reynolds stress ($\overline{w'w'}$), only the redistribution term is significant. The influence of one flow over another in the co-flow state was observed through the additional mean velocity and Reynolds normal stress found in the system compared to a standard plane Couette flow (pCf). Comparing the co-flow system with a conventional pCf system, the former exhibits greater vorticity, vortex stretching, and kinetic energy. A detailed analysis on the geometry and topology of flow structures was studied using flow invariants.

Published under an exclusive license by AIP Publishing. https://doi.org/10.1063/5.0107519

## I. INTRODUCTION

There exist numerous flow systems where turbulent and non-turbulent flows are adjacent to each other (e.g., jets, wakes, plumes, and boundary layers). Most studies often limit to studying the turbulent region, to understand turbulence dynamics of the flow system. However, this leaves an important aspect unattended—how does the flow transit from a turbulent to non-turbulent region? Relatively fewer studies approached this problem by studying the interface region,[1–5] which revealed some fundamental characteristics of the interface and the flow physics in the region. The interface is an extremely thin region with a thickness of order Taylor length scale,[6,7] comprised of two layers—the laminar superlayer or viscous superlayer (VSL) and the turbulent sub-layer (TSL). The theoretical and computational arguments about the existence of these layers were discussed by da Silva et al.[3] The observation of the viscous sub-layer (VSL) was first made by Taveira and da Silva.[8] Figure 1(a) shows a schematic of the three regions—turbulent, irrotational, and the interface. The figure further shows the two sub-layers of the interface—VSL and TSL. The VSL lies adjacent to the irrotational region, and the TSL lies adjacent to the turbulent region. The primary purpose of the existence of VSL is to induce vorticity, which is achieved through a process called viscous diffusion.[9] The genesis of vorticity in the VSL helps to smoothly match the vorticity from zero (in the irrotational region) to a finite value (in the TSL). The vorticity increases as we go into the interface region (from VSL to turbulent region, through TSL). The extremely small thickness of the interface results in sharp vorticity gradients in the region, which is proposed to be a characteristic feature of TNT flows of all classes (jets,[10–13] plumes,[14] wakes,[1] and boundary layers[5,15,16]).

A great amount of work was particularly done in the class of jets. In the work by Teixeira and da Silva,[10] adjacent isotropic turbulent and non-turbulent regions were studied analytically and computationally using Rapid Distortion Theory (RDT) and direct numerical simulation (DNS), respectively. In their work, da Silva and Pereira[11] have simulated a turbulent plane jet and studied the topology of turbulent structures and the dynamics of flow across the turbulent non-turbulent interface (TNTI) using invariants of velocity gradients, rate-of-strain, and rate-of-rotation. The same plane jet configuration was used by Taveira and da Silva to study the kinetic energy and perform budget analysis.[4] Breda and Buxton[17] have studied the TNTI region and compared the dynamics between a round jet and fractal jet. The study on TNTI dynamics in wakes behind a flat plate was done by Bisset et al.[1] The article elaborates on various global and local dynamics like turbulent kinetic energy, entrainment, etc., giving an overview of various aspects. The work done by Krug et al.[14] gave a holistic view on kinetic energy and entrainment dynamics in turbulent plumes. In the class of wall-bounded flows, TNT studies were largely made in the boundary layer flows. Chauhan et al.[5] studied the dynamics and entrainment process in boundary layer flows.

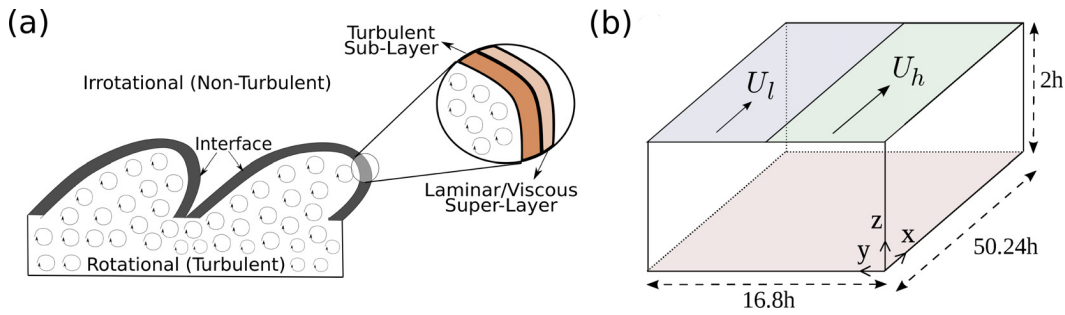

**FIG. 1.** (a) Schematic representation of turbulent/non-turbulent interface region and its sub-layers sandwiched between the turbulent and non-turbulent regions, (b) current flow configuration and the dimensions of computational flow system. x, y, and z represent the streamwise, spanwise, and wall-normal directions, respectively. The $U_h$ and $U_l$ represent the velocities of the two adjacent top plates.

Most of the studies are limited to only the open flows like jets and wakes, and the study of such adjacent turbulent/non-turbulent systems remains obscure in semi-confined and confined flows. Except the preliminary studies by Narasimhamurthy et al.[18,19] and Teja et al.,[20,21] no significant work has been done in this class of flows. Figure 1(b) shows the flow configuration used by Narasimhamurthy et al. and Teja et al. In the work by Teja et al.,[20] two plane Couette flows (pCfs) at different Reynolds numbers ($Re_h = 500$, $Re_l = 100$) are made to flow adjacent to each other in the same direction. Here, $Re_h$ and $Re_l$ correspond to the Reynolds number of flow sheared by plates moving at velocities $U_h$ and $U_l$. Reynolds number in the current flow is defined based on half channel height (h) and half the plate velocity. Thus, $Re_h = \frac{U_h h}{2\nu}$ and $Re_l = \frac{U_l h}{2\nu}$. Previous studies on pCf reported the critical Reynolds number to be approximately 500.[22,23] This was also confirmed by Teja et al.[20] Thus, the Reynolds number combination ($Re_h, Re_l$) = (500, 100) represents the co-flow of turbulent and laminar pCfs. Teja et al.[20] presented the kinematic analysis of adjacent turbulent/non-turbulent flows for different Reynolds number combinations ($Re_h, Re_l$) and Reynolds number ratios ($r = Re_h/Re_l$). However, several dynamical aspects like the effect of one flow over the other, energy budget analysis[24–26] of the flow system, vorticity dynamics,[24,27,28] and topology[29] still remain unexplored. This article is aimed toward extending the study of Teja et al.[20] using the same data to analyze further, studying the dynamics of unstable co-flowing plane Couette flow for a particular Reynolds number combination ($Re_h, Re_l$) = (500, 100). A qualitative visualization of the base flow (defined in Teja et al.[20]) and co-flow can be seen in Fig. 6.

It is important to understand the novelty of the current flow system relative to other systems studied so far. In the other flow systems (like jets, wakes, boundary-layers, etc.), the non-turbulent region is quiescent and unbounded, and, thus, the turbulence tends to freely diffuse into surroundings, leading to flow dynamics like the entrainment.[30] In the current system, the flow is forced to remain at a specific sub-critical Reynolds number ($Re_l = 100$) (and thus in viscous dominated region) through the top plate velocity ($U_l$). Any turbulence transported from the adjacent turbulent region is thereby subjected to viscous effects. This prohibits the turbulence entrainment. In addition, the entire system is semi-confined by walls (on the top and the bottom). One of the applications of the plane Couette flows[31] and the current flow configuration is the movement of ships in the sea. Gourley[32] considered the keel of a flat bottomed ship as the moving top plate of a plane Couette flow and the sea bed as the stationary bottom plate of the plane Couette flow. The fluid motion between them resembles the flow characteristics of the plane Couette flow. The current problem of co-flowing plane Couette flows replicates the scenario of ship-to-ship cargo transfer operation. During this process, a daughter vessel approaches the mother vessel. During the approach and departure of a daughter ship, the two vessels have different velocities, thus resembling the scenario of two co-flowing plane Couette flows with different top plate velocities.

Though the system is unstable and periodic in nature, the objective of the current study is to understand the influence of a turbulent flow on the adjacent co-flowing non-turbulent flow. In order to understand the large scale dynamics of the shear-layer, the analysis of mean quantities is considered. Upon averaging, the mean quantities have no wave characteristics (see Fig. 2). It is, hence, considered to neglect the wave analysis of the flow in the current study. Thus, the interface is statistically a straight line at $y \approx 8.4h$. However, this does not imply that the wave has no impact on flow variables, but the wave effect has been observed through the higher order statistical quantities. While including the wave dynamics would make the problem more comprehensive and interesting, however, this is beyond the scope of current investigation.

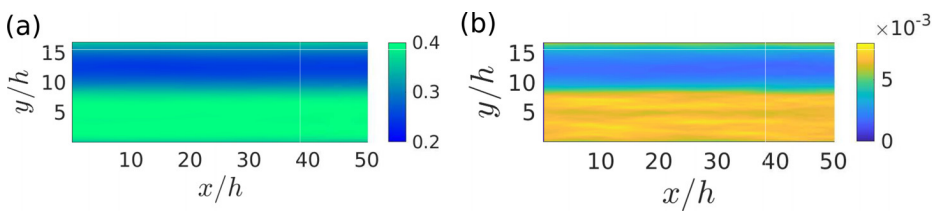

**FIG. 2.** Contours of (a) mean streamwise velocity, (b) turbulent kinetic energy ($\bar{k}$) at $z = 1h$. Here, $\bar{k}$ is defined as $\bar{k} = \frac{1}{2}(\overline{u'u'} + \overline{v'v'} + \overline{w'w'})$. The x and y represent the streamwise and spanwise directions, respectively, and are normalized by h.

Since the interface is fixed at $y \approx 8.4h$, for a clear distinction between the turbulent and non-turbulent regions, a modified $y$-axis ($y_I$) is used for analysis of the mean quantities. $y_I$ is obtained by subtracting the interface location from $y$, i.e., ($y_I = 8.4h - y$). The negative and positive values of $y_I$ correspond to the turbulent and non-turbulent regions, respectively.

The article introduces the computational code, the domain, and grid details in Sec. II. Subsections III A and III A 2 discuss the vortex dynamics of the flow problem. A comparison between the base flow (plane Couette flow) and the co-flow is made wherever relevant. Subsection III B uses the invariant analysis to understand the flow topology observed in Secs. III A and III A 2. Subsection III C explains the effect of turbulent pCf on the adjacent non-turbulent pCf through mean and turbulent quantities. The analysis of turbulence transportation through the budget analysis of the turbulent normal stress components is explained in Subsection III D. Section IV reports the summary of the findings.

## II. NUMERICAL METHODOLOGY

In the present direct numerical simulation (DNS) study, the complete Navier–Stokes equation for an isothermal and incompressible flow was solved using a parallel finite volume code, MGLET.[33] The solver adopts staggered Cartesian mesh. The spatial discretization of convective and diffusive fluxes is performed using a second-order accurate central differencing method. The momentum equations are marched in time by a fractional time-stepping method using a second-order accurate Adams–Bashforth method. Finally, the Poisson equation is numerically solved by a multi-grid technique based on point-wise velocity-pressure iterations.

In this study, a computational domain of lengths $50.24h$, $16.80h$, and $2h$ along the streamwise ($x$), spanwise ($y$), and wall-normal ($z$) directions, respectively, is taken. The same domain was used in the previous studies of Narasimhamurthy et al.[18,19] and Teja et al.[20,21] Figure 1(b) shows the computational domain and its orientation. The boundary conditions for the top and the bottom walls were given as no-slip, while the other sides were given the periodic boundary condition. In a physical sense, this configuration corresponds to a spanwise arrangement of multiple plane Couette flows with alternating high and low Reynolds numbers. Thus, the present domain in Fig. 1(b) shows two interface zones—one at the start or end of the domain ($y = 0h$ or $16.8h$), and one at the center of the domain ($y = 8.4h$) along the spanwise direction.

A structured grid consisting of 256, 256, and 64 grid points along the $x$, $y$, and $z$ directions, respectively, is taken. While the grid is uniformly spaced along the $x$ and $y$ directions, it is stretched along the wall-normal $z$-direction. The mesh is kept fine close to the walls, and it is stretched as we move toward the channel center. For the base state Reynolds number (as defined in Teja et al.[20]) $Re = 500$, the grid resolution (in terms of viscous wall units) along the $x$ and $y$ directions is 7.34 and 2.46, respectively, and it varies from 0.9 at the walls to 1.48 at the channel center along the $z$-direction. Figures 3(a) and 3(c) show the profiles of streamwise and spanwise averaged Taylor and Kolmogorov length scales for the base state Reynolds number (see Teja et al.[20]) of 500, respectively. The Taylor length scale is defined as $\lambda = \sqrt{\frac{\langle u'u' \rangle}{\langle \frac{\partial u'}{\partial x} \frac{\partial u'}{\partial x} \rangle}}$.[34]

Figures 3(a) and 3(b) shows the variation of Taylor length scale ($\lambda$) and Reynolds number (calculated based on $\lambda$ using the formula $Re_\lambda = \frac{v_{rms}\lambda}{\nu}$) across the wall-normal direction, respectively. Figure 3(b) shows the flow Reynolds numbers calculated based on length scales that are most significant. The Kolmogorov length scale is defined as $\eta = \sqrt[4]{\frac{\nu^3}{\epsilon}}$.[4] The value of $\eta$ in the current case varies approximately between 0.043 and 0.048 in the domain [see Fig. 3(c)]. A detailed study on the adequacy of the current grid and domain has been explained in Teja et al.[20]

## III. RESULTS
### A. Vorticity

One of the most effective ways to distinguish the turbulent and non-turbulent regions and the transition between the regions is by using the enstrophy. Enstrophy is defined as the dot product of instantaneous vorticities ($\omega_i \omega_i$). Figure 4 shows the spanwise variation of enstrophy. The high values of enstrophy between $y \approx [0, 8]$ depict that the region is turbulent. The approximate zero magnitude of enstrophy between $y \approx [9, 15]$ denotes that the region is non-turbulent. The interface locations ($y \approx 8.4h, y \approx 16.8h$) exhibit a sharp jump signifying the change in the region. Such a jump at turbulent/non-turbulent interface was also observed in jets, wakes, and boundary layers.[3,15,17,35] This indicates that the jump is characteristic feature of TNTI irrespective of class of flow.

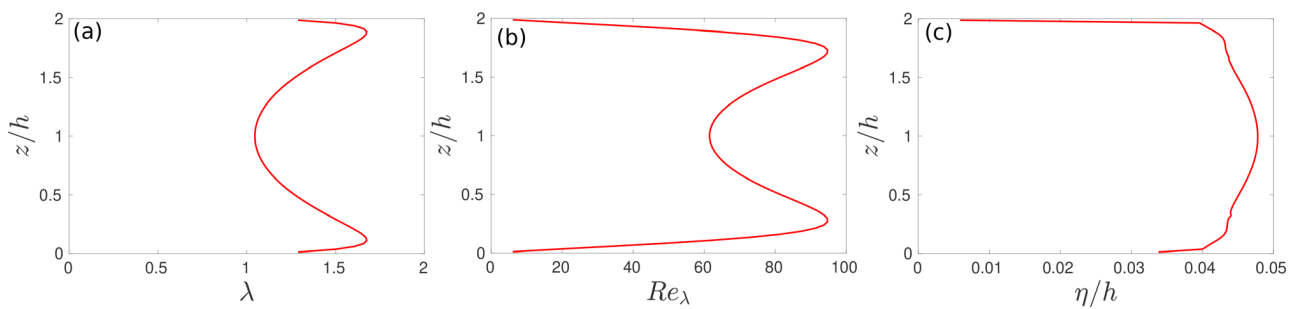

**FIG. 3.** Profiles of (a) Taylor length scale ($\lambda$), (b) Taylor length scale based Reynolds number ($Re_\lambda$), and (c) Kolmogorov length scale ($\eta$) for the base state Reynolds number 500 in the plane Couette flow. Data here stem from both time averaging and spatial averaging along the homogeneous streamwise ($x$) and spanwise ($y$) directions.

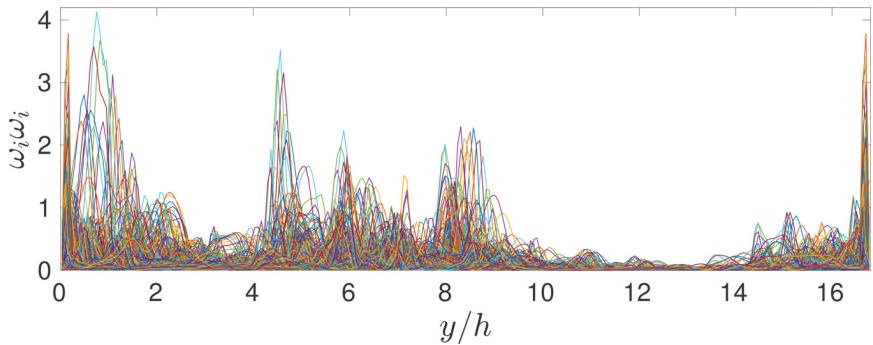

FIG. 4. Variation of instantaneous enstrophy ($\omega_i\omega_i$) along the spanwise ($y_l$) direction. The variation is plotted for all the streamwise ($x$) locations (each location represented by a different colored line) and at mid wall-normal plane ($z = 1h$). The enstrophy is normalized by $U_h/h$, and $x$ and $y$ are normalized by $h$.

A non-turbulent region is characterized by zero vorticity (and hence zero enstrophy). However, Fig. 4 shows a near zero enstrophy in the non-turbulent region. We hypothesize that such a non-zero enstrophy could be due to walls (studied in Sec. III A 1) and/or due to turbulence transportation (explored later in the article).

### 1. Variation of enstrophy and wave characteristics along wall-normal direction

To understand the origins of enstrophy in the non-turbulent region, instantaneous enstrophy contours plotted at different wall-normal locations in Fig. 5. We notice that the magnitude of enstrophy in the non-turbulent region is large close to the bottom wall [see Fig. 5(e)] and decreases as one moves toward the top wall [see Fig. 5(a)]. This suggests that the enstrophy could have possibly originated at the stationary bottom wall $z = 0$ due to irrotational/potential vorticity.

Figure 5 also shows the variation in wave characteristics along $z$. This depicts the bilateral nature of the flow instability. The influence of bilateral nature of the flow problem can also be seen in turbulent kinetic energy (see Fig. 17 in Sec. III D 4). However, a detailed analysis of this feature is not in the scope of current article and could be interesting for a future research.

### 2. Comparison of vorticity between base flow and present co-flow

A qualitative comparison of vorticity is made using the three-dimensional iso-contours to understand how the co-flow system differs from the base flow (standard pCf). Figures 6(a) and 6(b) compare the iso-contours of streamwise component of vorticity ($\omega_x$) between the base flow (left) and co-flow (right). The positive values of $\omega_x$ are clockwise rotating vortices and are colored in green. The structures with a negative value of $\omega_x$ are colored red and represent the counter-clockwise rotating vortices. While the structures are small and dense in the base flow, the structures are bulky and sparse in the co-flow. The wide and long structure of vortices in co-flow could be due to the shear effect.

Similarly, Figs. 6(c) and 6(d) compare the iso-contours of spanwise component of vorticity ($\omega_y$) between the base flow (left) and the co-flow (right). The structures with positive values of $\omega_y$ rotating in the clockwise direction are colored in yellow. The blue-colored structures are the vortices with counterclockwise rotation and have negative values of $\omega_y$. While the structures in base flow are dense and evenly distributed, the structures in co-flow are sparsely distributed. Furthermore, we observe that the turbulent region is dominated by clockwise vortices. While the counterclockwise vortices tend to appear upon decreasing the iso-contour value, they are significantly less in number. Thus, the turbulent flow predominantly contains the clockwise vortices.

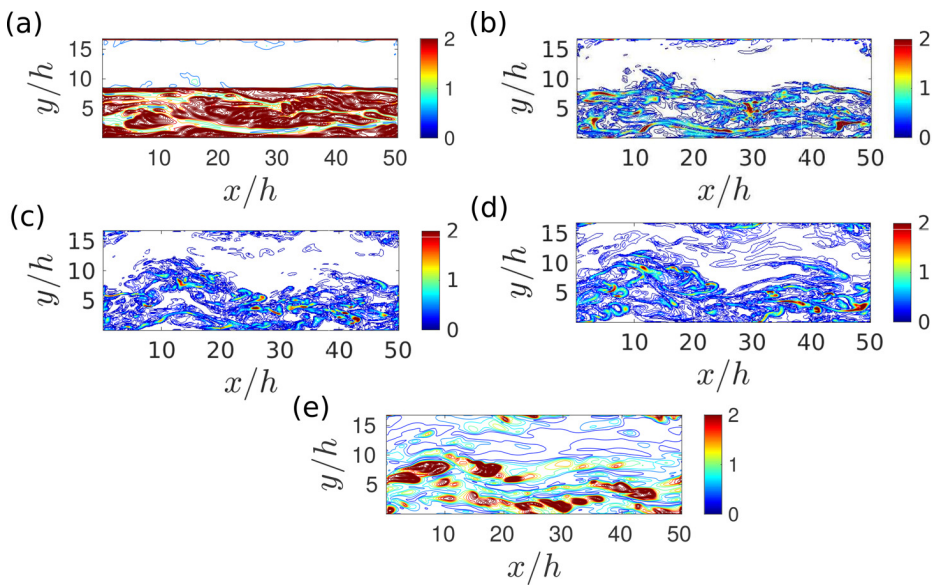

FIG. 5. Instantaneous enstrophy ($\omega_i\omega_i$) contours plotted at different wall-normal ($z$) locations: (a) $z = 1.9h$; (b) $z = 1.5h$; (c) $z = 1h$; (d) $z = 0.5h$; and (e) $z = 0.1h$. The $x$ and $y$ represent the streamwise and spanwise directions, respectively.

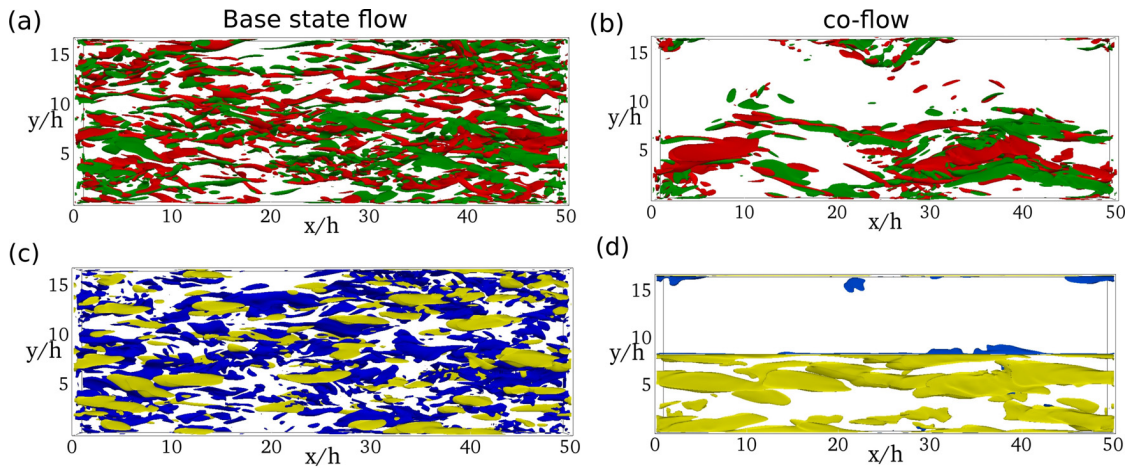

**FIG. 6.** Three-dimensional iso-contours showing: (a) and (b) streamwise component of vorticity ($\omega_x$) and (c) and (d) spanwise component of vorticity ($\omega_y$). The contours on the left and right show the base state (pCf) and co-flow state, respectively. The flow Reynolds number of the base flow is 500, and the Reynolds number combination of co-flow is (500, 100). The red and blue colored structures indicate the negative iso-contour value, and green and yellow colored structures indicate the positive iso-contour value.

We see that the vortical structures in Fig. 6 are in the form of sheets. This is particularly clear in the co-flow [Figs. 6(b) and 6(d)]. A more detailed analysis on this flow topology and geometry is done using invariant analysis in Sec. III B.

*a. Vorticity dynamics.* Due to the co-flow of pCfs at different Reynolds numbers, it is possible that the shear could influence the vortices, of particular interest here is the straining of vortices. This is studied in the base state and co-flow states using the vortex stretching. The origin of this term from the vorticity equation can be seen in Appendix A.

Vortex stretching is the elongation of vortices in the fluid due to the mean flow. The term $\omega_j \frac{\partial U_i}{\partial x_j}$ in the vorticity equation [Eq. (A1)] numerically represents this phenomenon. The elongation of vortices is accompanied by increase in the strength of the vortex.

### 3. Comparison of vortex stretching between base flow and co-flow

Figures 7(a) and 7(b) show the three-dimensional iso-contours of streamwise component of vortex stretching ($\omega_j \frac{\partial U}{\partial x_j}$) between a standard pCf (left) and a co-flowing turbulent and non-turbulent flows (right). A comparison of turbulent structures between the streamwise vorticity [Figs. 6(a) and 6(b)] and the streamwise vortex stretching [Figs. 7(a) and 7(b)] shows that the vortical structures are thinner and elongated.

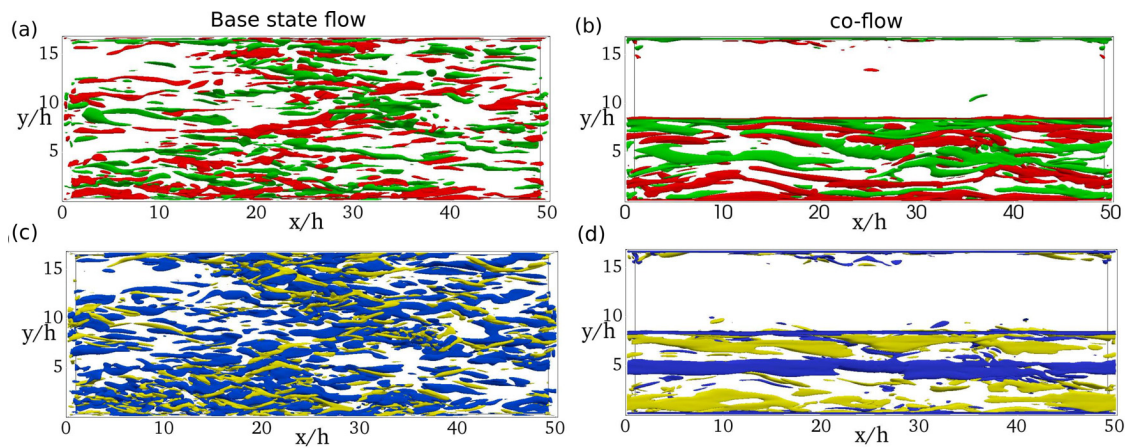

**FIG. 7.** Three-dimensional iso-contours showing the (a) and (b) streamwise component of vortex stretching ($\omega_j \frac{\partial U}{\partial x_j}$) and (c) and (d) spanwise component of vortex stretching ($\omega_j \frac{\partial V}{\partial x_j}$). The contours on the left and right show the base state (pCf) and co-flow state, respectively. The flow Reynolds number of the base flow is 500, and the Reynolds number combination of co-flow is (500, 100). The red and blue colored structures indicate the negative iso-contour value, and green and yellow colored structures indicate the positive iso-contour value.

A similar effect can be observed in co-flow, except that the structures are more elongated than base flow. This demonstrates the straining of vortical structures.

Similarly, Figs. 7(c) and 7(d) show the spanwise component of vortex stretching ($\omega_j \frac{\partial V}{\partial x_j}$). Unlike Figs. 7(a) and 7(b), the contours barely show stretching along the spanwise direction [compare Figs. 7(c) and 7(d) with 6(c) and 6(d)]. This is due to the small magnitude of $\frac{\partial V}{\partial x_j}$, which barely stretches the vortex $\omega_j$. On the contrary, they are much thinner than structures in Figs. 6(c) and 6(d). This tells that the thinning due to streamwise elongation is more dominant than the elongation due to spanwise velocity.

### B. Invariants analysis

We have noticed that the vortical structures showed a sheet-like structure in Fig. 6. To further understand the topology, we use the invariants of velocity gradient tensor ($U_{i,j}$) and its components—the rate of strain ($S_{ij}$) and the rate of rotation ($\Omega_{ij}$) tensors. This helps to understand not only the topology but also the geometry and dynamics at local level. The reader is referred to Appendix B for a theoretical understanding of the invariant analysis and how it helps to determine the topology, dynamics on flow structures. The invariants are robust tools for the analysis since they are intrinsic part of the flow system and independent of the coordinate-axis.[36] Furthermore, the invariant maps aid in understanding the topological and geometric details that are abstruse in contour plots.

#### 1. Quantifying the invariants

Since the invariants in the current study are calculated from the mean quantities, they are homogeneous along $x$ and can be averaged along the same. Figure 8 shows the spanwise variation of various invariants plotted at mid-channel location ($z = 1h$). We notice that $\overline{Q_\Omega}$ and $\overline{Q_s}$ are the most dominant quantities representing that the flow is characterized by high enstrophy and relatively dominated by the dissipation process. A plot comparing the three invariants ($\overline{Q}, \overline{R}, \overline{R_s}$) shows that $\overline{Q}$ has relatively significant magnitude, but other quantities can be neglected.

#### 2. Comparison of invariant maps between base flow and co-flow

In order to understand how the geometry, topology, and dynamics in co-flow differ from conventional pCf, invariant maps obtained from the mean quantities are compared against each other. The left [(a)–(c)] and right [(d)–(f)] panels of Fig. 9 show the invariant maps of turbulent pCf (base flow) and co-flowing pCfs, respectively.

The blue and red colored triangles in Figs. 9(d)–9(f) represent the non-turbulent and turbulent regions, respectively. Clearly, except the $(\overline{Q_\Omega}, \overline{Q_s})$ invariant map, the other invariant maps differ significantly between the base state flow and co-flow state. The $(\overline{R}, \overline{Q})$ map for the base state [Fig. 9(a)] shows that $\overline{Q}$ is predominantly positive, and, hence, it is dominated by vortical structures. The value of $\overline{R}$ being both positive and negative, and the vortical structures undergo both elongation and compression. A greater number of points with positive $\overline{R}$ signify that a greater number of vortical structures undergo compression than elongation. In contrast, the co-flow has both positive and negative values of $\overline{Q}$ and hence has both vortical and dissipative structures. Unlike the base state, a greater proportion of points in co-flow have negative $\overline{R}$, implying that more structures undergo elongation than compression. This is in agreement with the observation in Sec. III A 2 and could be due to shearing effect present in co-flow. The $(\overline{Q_\Omega}, -\overline{Q_s})$ maps show that both the base state and co-flow state have vortex sheet structures, i.e., the viscous dissipation and enstrophy are equally dominant.

To understand how the flow dynamics, geometry, and topological features in co-flow state change with wall normal location, the invariant maps are plotted at different $z$ for the co-flow state. Figures 10(a)–10(c) show the invariant maps close to stationary bottom wall ($z = 0.2h$), and Figs. 10(d)–10(f) show maps close to the moving top wall ($z = 1.8h$). The invariant maps at channel core ($z = 1h$) can be seen from Figs. 9(d)–9(f). The $(\overline{R}, \overline{Q})$ maps show the presence of both vortical and dissipation structures close to walls. However, in the channel core, we only find the vortical structures. The dominant dissipation happening at walls could be responsible for the appearance of dissipative structures at $z = 0.2h$ and $z = 1.8h$. The presence and absence of vortical structures in the non-turbulent region at different heights (seen in Fig. 5) can also be seen from $(\overline{R}, \overline{Q})$ invariant maps. In Fig. 10(d), the non-turbulent region has no positive $Q$, thus signifying the absence of vortical structures in the non-turbulent region. The $(\overline{R_s}, \overline{Q_s})$ invariant maps at different wall-normal location show that the fluid elements in the turbulent region undergo only expansion close to the bottom wall and predominantly compression in the channel core. At the top wall, we see both the compression and expansion of vortical structures. The elements in the laminar region, however, exhibit a monotonous decrease in deformation (compression) as we move from bottom wall to top wall. The $(\overline{Q_\omega}, \overline{Q_s})$ maps show that the vortical structures remain vortex sheets at all wall-normal locations.

### C. First-order and second-order statistics

The influence of one flow over another is understood through the mean quantities of co-flow, which are plotted at various spanwise locations, and the profile is compared with the plane Couette flow (pCf). The mean streamwise velocity and streamwise Reynolds normal stress profiles for the base flow and co-flow are plotted in Fig. 11. A comparison of mean streamwise velocity for the base flow [Fig. 11(a)] and turbulent co-flow [Fig. 11(c)] shows a similar "S" shaped profile in the turbulent region. However, it is interesting to see that the mean velocity of the non-turbulent region of co-flow shows a curved profile. Recall that the mean streamwise velocity of a laminar pCf has a linear variation.[37] A comparison of mean velocity profiles between the conventional pCf and co-flowing pCf in the laminar regime is plotted in Fig. 12. The figure shows the additional momentum existing in the co-flow system. Furthermore, in Fig. 11(c), one can see that the mean velocity is much higher at the boundary of the non-turbulent region than the core (also shown in Fig. 12). This continuous variation in mean velocity along $y$ leads to shearing across multiple layers in the spanwise direction. The enhanced momentum in the non-turbulent region could possibly due to the adjacent high Reynolds number flow.

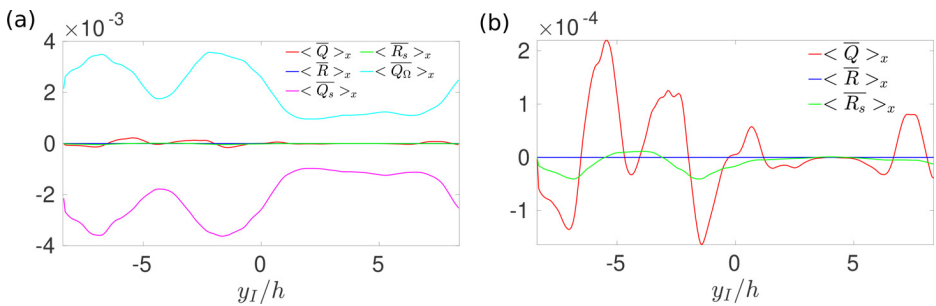

FIG. 8. Spanwise variation of invariants: (a) $\langle \overline{Q} \rangle_x, \langle \overline{R} \rangle_x, \langle \overline{Q_s} \rangle_x, \langle \overline{R_s} \rangle_x, \langle \overline{Q_\Omega} \rangle_x$; (b) $\langle \overline{Q} \rangle_x, \langle \overline{R} \rangle_x, \langle \overline{R_s} \rangle_x$. The invariants are plotted based on mean statistics and are streamwise averaged. The figures are plotted at mid z-plane ($z = 1h$).

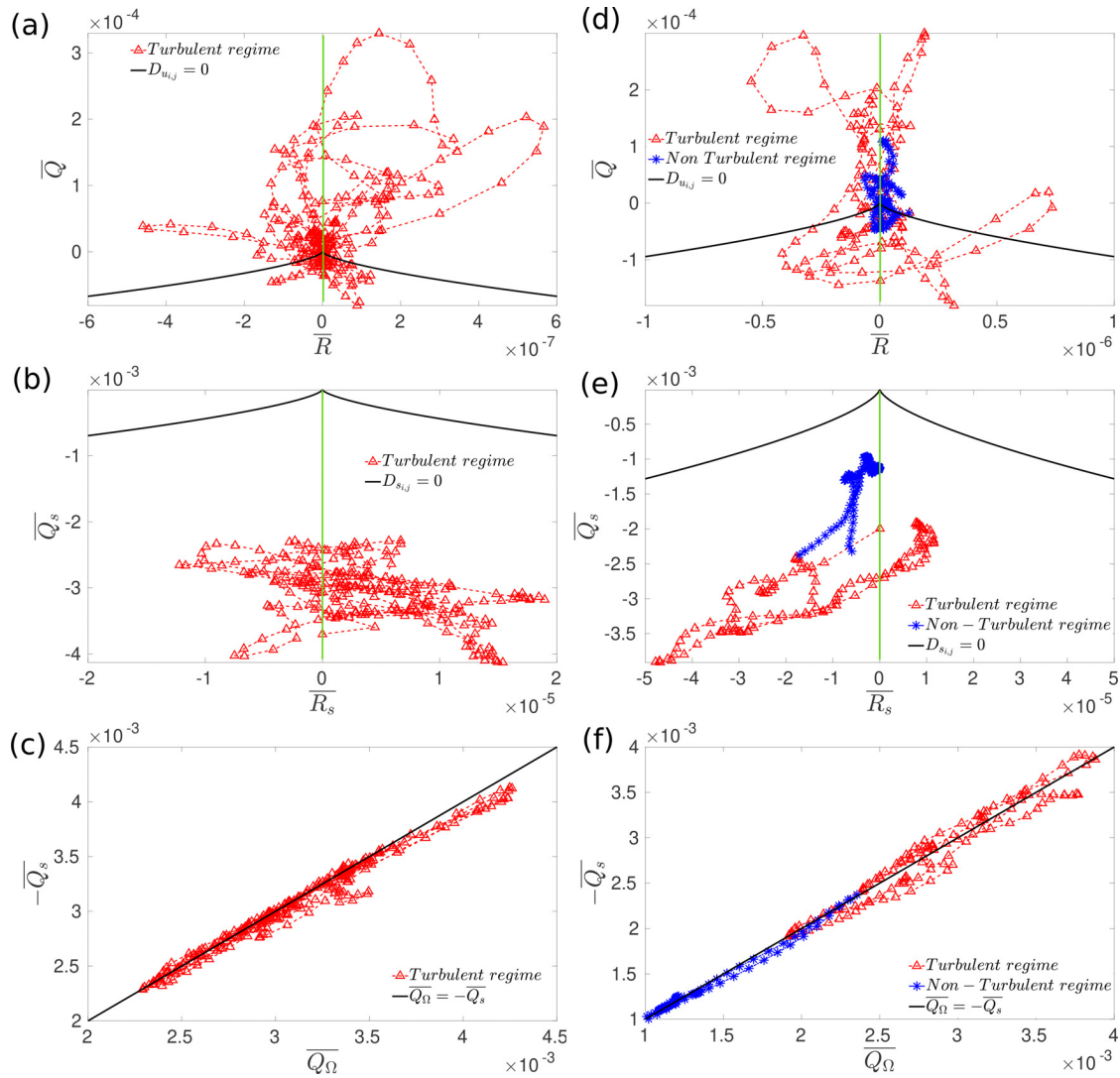

**FIG. 9.** Comparison of $(\bar{R}, \bar{Q})$, $(\bar{R_s}, \bar{Q_s})$, and $(\bar{Q_\Omega}, \bar{Q_s})$ invariant maps between base flow and co-flow. (a)–(c) show the three invariant maps for the base flow (pCf at Re 500), and (d)–(f) show the three invariant maps for the co-flowing pCfs. The figures are plotted along the spanwise direction (each triangle representing a spanwise grid-point) at mid streamwise and wall-normal plane ($x = 25.12h, z = 1h$).

Figure 11(b) shows the wall-normal variation of $\overline{u'u'}$ for the base flow (turbulent pCf at Reynolds number 500). The profile of $\overline{u'u'}$ for a turbulent pCf is symmetric about channel centerline ($z = 1h$). However, this symmetry is not observed in the turbulent region of co-flowing pCf, and the profile is more skewed toward the top plate [see Fig. 11(d)], i.e., the magnitude of $\overline{u'u'}$ is higher at the top wall and lower at the bottom wall. Furthermore, Fig. 11(d) also shows the presence of $\overline{u'u'}$ in the non-turbulent zone. Such an observation was also made in previous studies like Xavier et al.[38] and Philips.[39] This implies that the turbulence is being transported from the turbulent region to the non-turbulent region. Similarly, the profile of $\overline{v'v'}$ and $\overline{w'w'}$ showed non-zero magnitudes in the non-turbulent region. The presence of Reynolds normal stresses in the non-turbulent region demands further investigation, and the budget analysis of Reynolds normal stress terms is done toward this end.

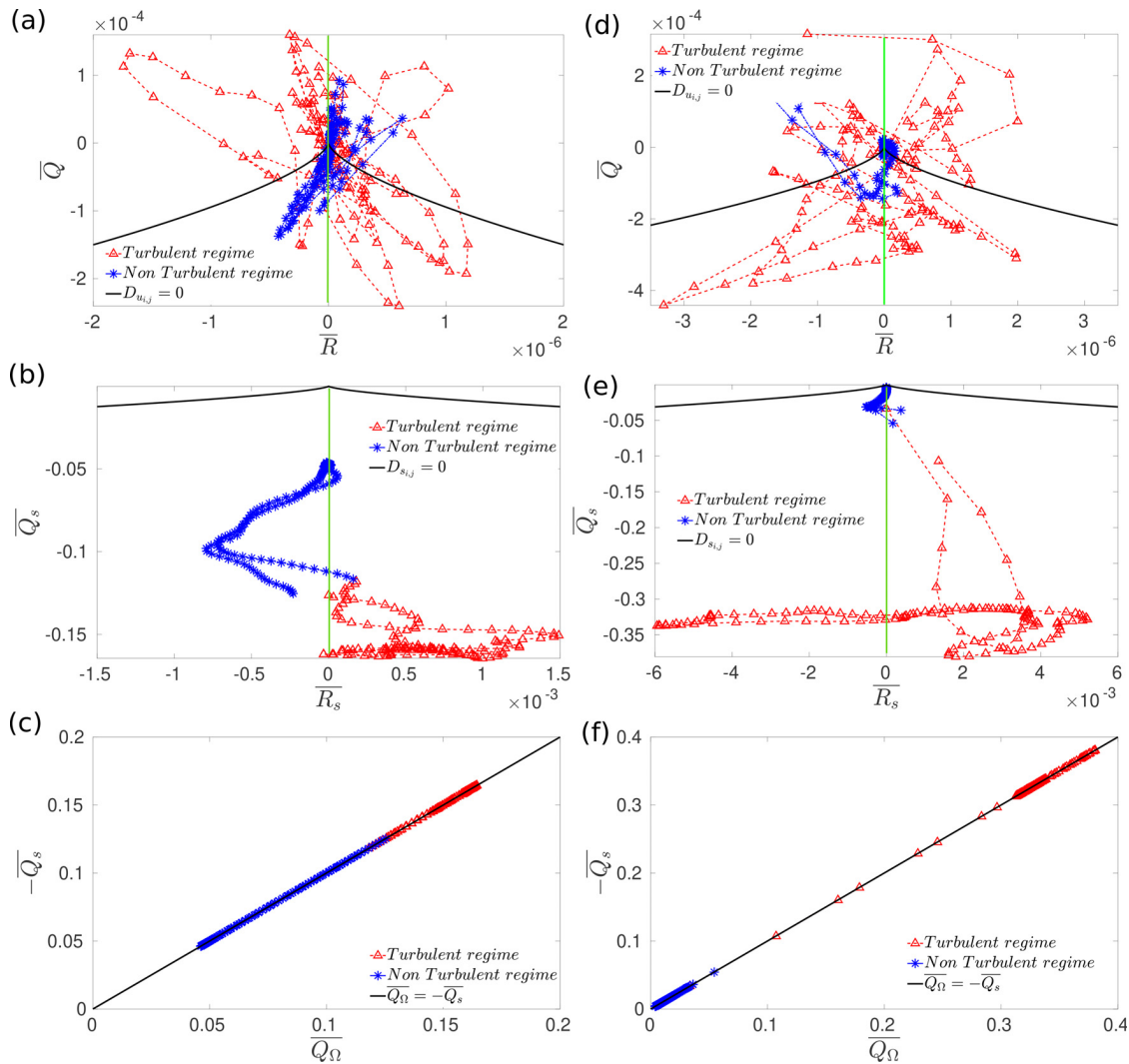

**FIG. 10.** Comparison of invariants for co-flowing plane Couette flows at different wall-normal planes. (a)–(c) $(\overline{R}, \overline{Q})$, $(\overline{R_s}, \overline{Q_s})$, and $(\overline{Q_\Omega}, \overline{Q_s})$ invariant maps at $z \approx 0.2h$ (bottom plate); (d)–(f) represents the same maps at $z \approx 1.8h$ (top plate).

### D. Budget analysis

The reader is referred to Appendix C to see the derivation of the turbulent budget equation. This section focuses on budget analysis of the three Reynolds normal stress terms. The budget equation for each term is obtained by inserting desired values of $i$ and $k$ in Eq. (C1). Since our current interest is to study the dynamics across the interface, we plot the variation of each term along the spanwise axis. Each term of the budget equation being homogeneous along $x$ is averaged in that direction. The resulting planar data are taken, and budget analysis is done at a particular wall-normal location, here at $z = 1h$.

#### 1. $\overline{u'u'}$ budget

The budget equation for $\overline{u'u'}$ is obtained by taking $i = k = 1$ in Eq. (C1). Since we deal with mean quantities, the temporal term becomes zero, and the spatial derivative along $x$ becomes zero. Based on the above conditions, the $\overline{u'u'}$ budget equation simplifies to

$$0 = -\bar{U}_j \frac{\partial \overline{u'u'}}{\partial x_j} - 2\left[\overline{u'u'_j}\frac{\partial \bar{U}}{\partial x_j}\right] - \frac{\partial \overline{u'u'_j u'}}{\partial x_j} - \frac{2}{\rho}\frac{\partial \overline{p'u'}}{\partial x} + \nu \frac{\partial^2 \overline{u'u'}}{\partial x_j^2}$$
$$+ 2\left[\overline{\frac{p'}{\rho}\frac{\partial u'}{\partial x}}\right] - 2\nu \overline{\frac{\partial u'}{\partial x_j}\frac{\partial u'}{\partial x_j}}, \quad j \in [2,3] \forall j \in \mathbf{N}. \quad (1)$$

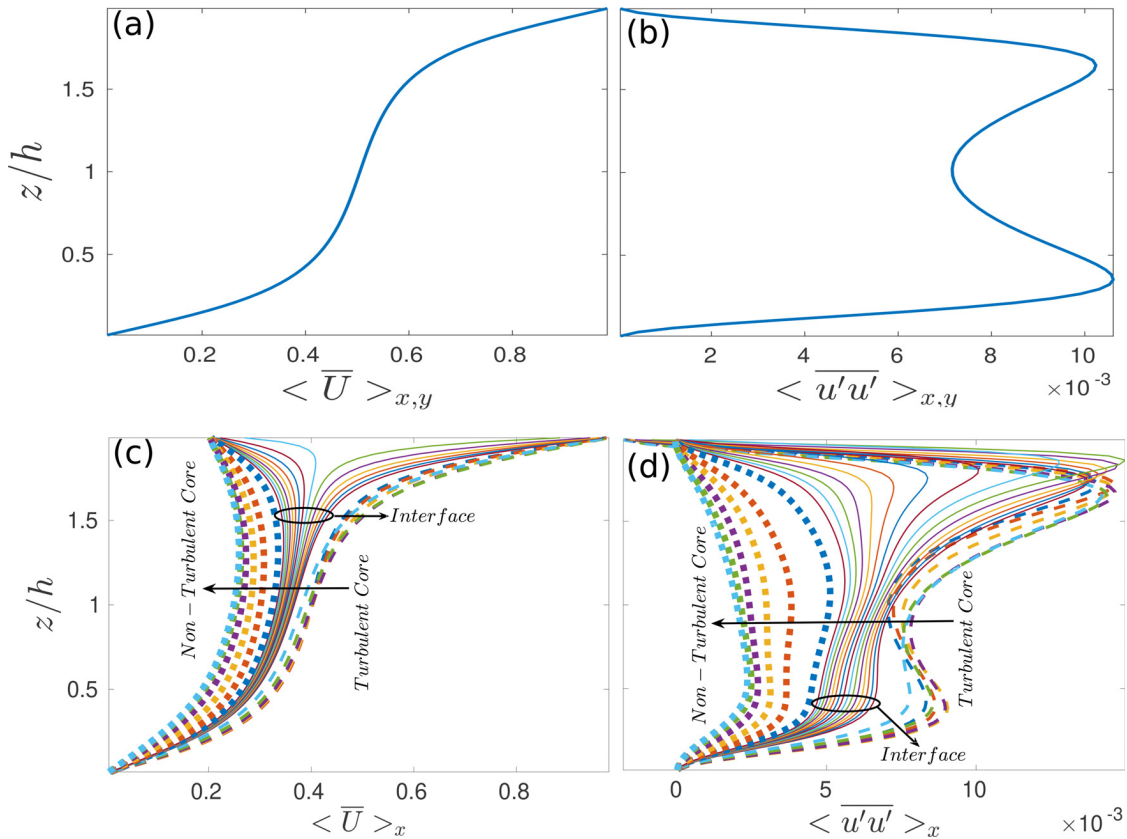

**FIG. 11.** Wall-normal (z) variation of (a) and (c) mean streamwise velocity and (b) and (d) Reynolds normal stress term. The top panel (a) and (b) shows the results for base state flow ($Re = 500$). The bottom panel (c) and (d) shows results for co-flow ($Re_h, Re_l$) = (500, 100). Flow variables are spatially averaged in homogeneous directions—for the base state along $x, y$ and for the co-flow state along the $x$ direction. Quantities are plotted for different spanwise (y) locations. The turbulent core, non-turbulent core, and interface regions are shown clearly. The dashed lines correspond to the turbulent core region, the solid lines correspond to the interface, and the dotted lines correspond to the non-turbulent region. The overline represents the temporal averaging, and subscript represents the spatial averaging directions.

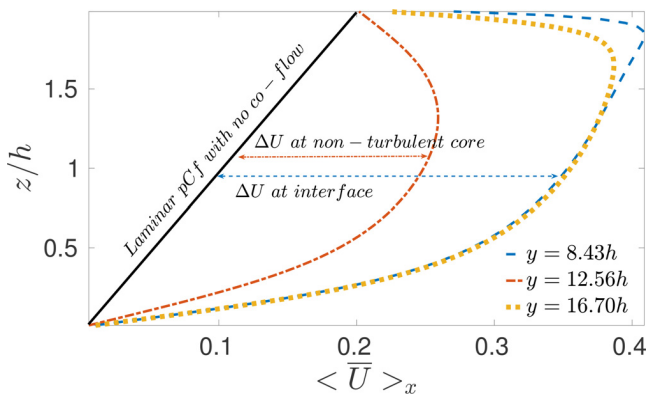

**FIG. 12.** Demonstration of momentum difference between an ideal laminar flow and the present non-turbulent region of co-flow. The dashed dot line shows profile at the non-turbulent core region ($y = 12.56h$), and dashed and dotted lines represent the profiles at interface locations ($y = 8.43h, y = 16.7h$).

Figure 13(a) shows the variation of each term with $y_I$. We see that the convection, viscous diffusion, and pressure diffusion (PD) terms are approximately zero all along the $y_I$. A closer look at the variation of the three terms along $y_I$ can be observed from Fig. 13(b). We see that the viscous diffusion and the convection term vary at an order less compared to the other terms. However, the pressure diffusion varies only at an order of $10^{-7}$ and, hence, is insignificant. The other terms—production, dissipation, and redistribution, which are significant in the turbulent region, undergo a "U" shaped variation in the non-turbulent region, decreasing gradually as they move from turbulent to the non-turbulent region (at the interface $y_I = 0$) and increasing gradually as we move from non-turbulent to turbulent regions (at the interface $y_I = 8.4h$). Note that the terms do not reach absolute zero anywhere in the non-turbulent region. The other term, turbulent diffusion (TD), which is significant in the turbulent region, drops to approximate zero right at the interface $y_I = 0$. The turbulent diffusion does not remain uniform all along the non-turbulent region but raises to a maximum value gradually till $y_I \approx 4.2h$ (i.e., half the non-turbulent region) and then decreases till the other interface at $y_I = 8.4h$. This variation can be seen from Fig. 13(b). Thus, the analysis suggests that the primary reason for the existence of $\overline{u'u'}$ in the non-turbulent region could be the production, which is due to the non-zero gradient of mean streamwise velocity.

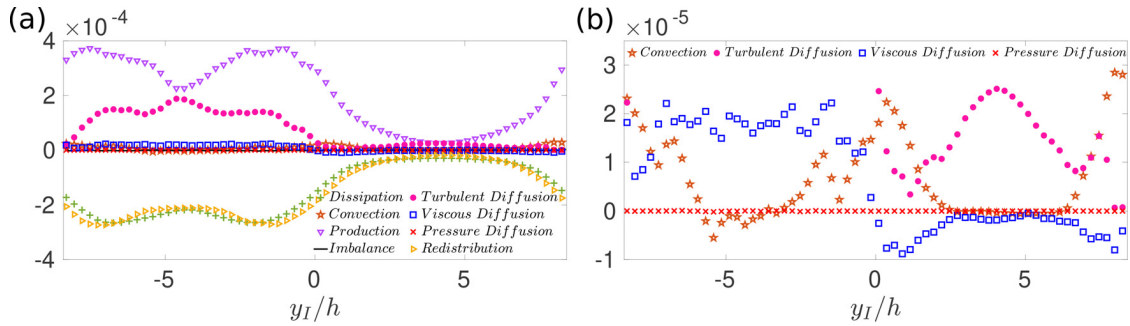

**FIG. 13.** Spanwise variation of (a) different budget terms and (b) convection, viscous diffusion, and pressure diffusion terms in the budget analysis of Reynolds stress term $\langle \overline{u'u'} \rangle_x$ for the Reynolds number combination (500, 100). The over-line and subscript represent the temporal and spatial *x* averaging. A modified *y*-axis (represented as $y_I$) is used here to differentiate turbulent and non-turbulent regions. Negative $y_I$ represents the turbulent region, and positive $y_I$ represents the non-turbulent region.

The process of turbulence generation can be understood from the production term (*P*), which is given by the formula

$$\mathcal{P}_{u'u'} = -2 \left[ \overline{u'u'_j} \frac{\partial \bar{U}}{\partial x_j} \right]. \tag{2}$$

Despite the negative sign before the term, we see that the term is positive [see Fig. 13(a)], which implies that $\overline{u'u'_j}$ and $\frac{\partial \bar{U}}{\partial x_j}$ are negatively correlated, i.e., turbulence is being produced at the expense of energy from the mean velocity gradient. With the pressure diffusion being insignificant and viscous diffusion contributing to a very small extent, the transportation of the turbulence is predominantly due to turbulent diffusion. This is true for the whole turbulent flow field and the non-turbulent core region. However, inside the non-turbulent region, close to the interface regions ($y_I \gtrsim 0, y_I \lesssim 8.4h$), we see that viscous diffusion is as much important as the turbulent diffusion [see Fig. 13(b)] and is negatively correlated. The turbulent diffusion (*TD*) is defined as follows:

$$\mathcal{TD}_{u'u'} = -\frac{\partial \overline{u'u'_j u'}}{\partial x_j}. \tag{3}$$

The positive and negative values of turbulence diffusion and viscous diffusion terms represent transportation of turbulence from non-turbulent to turbulent, and from turbulent to non-turbulent region, respectively. The redistribution and dissipation play an equally important role, i.e., while about half of the total available turbulent energy in the $\overline{u'u'}$ is redistributed into other directions, the other half is dissipated. The redistribution or pressure-strain rate (*PSR*) and the dissipation (*D*) terms are given by

$$\mathcal{PSR}_{u'u'} = 2 \left[ \frac{\overline{p'}}{\rho} \frac{\partial u'}{\partial x} \right]; \mathcal{D}_{u'u'} = 2\nu \frac{\overline{\partial u'_i}}{\partial x_j} \frac{\partial u'_i}{\partial x_j}. \tag{4}$$

With these terms playing important terms relatively to others, it is reasonable to assume that the energy balance in the turbulent region can be written in the form

$$\mathcal{P}_{u'u'} + \mathcal{TD}_{u'u'} + \mathcal{D}_{u'u'} + \mathcal{PSR}_{u'u'} \approx 0. \tag{5}$$

In the non-turbulent regime, significant mean velocity acts as a source for the development of turbulence. It is interesting to note that the turbulence on the non-turbulent side is mainly due to production than diffusion. The turbulence diffusion drops rapidly to near zero value close to the interface, and the viscous diffusion continues to remain insignificant, making the role of diffusion in transporting the turbulence to the non-turbulent side negligible. However, the production term drops gradually, and we see that the term is still significant in the non-turbulent region. The turbulence being produced is transported and destructed through the redistribution and dissipation processes, respectively. As we move along the spanwise direction toward the other interface at $y_I = 8.4h$, the production again raises after a minimum in the non-turbulent core. The redistribution and dissipation come back into play. The turbulent diffusion remains insignificant in the whole non-turbulent region, which starts to grow only after the interface $y_I = 8.4h$. Hence, the energy balance here becomes

$$\mathcal{P}_{u'u'} + \mathcal{D}_{u'u'} + \mathcal{PSR}_{u'u'} \approx 0. \tag{6}$$

### 2. $\overline{v'v'}$ budget

To obtain the budget equation for $\overline{v'v'}$, we take $i = k = 2$. The simplifications applied above are implemented here which result in an equation as follows:

$$0 = -\bar{U}_j \frac{\partial \overline{v'v'}}{\partial x_j} - 2 \left[ \overline{v'u'_j} \frac{\partial \bar{V}}{\partial x_j} \right] - \frac{\partial \overline{v'u'_j v'}}{\partial x_j} - \frac{2}{\rho} \frac{\partial \overline{p'v'}}{\partial y} + \nu \frac{\partial^2 \overline{v'v'}}{\partial x_j^2}$$
$$+ 2 \left[ \frac{\overline{p'}}{\rho} \frac{\partial v'}{\partial y} \right] - 2\nu \frac{\overline{\partial v'}}{\partial x_j} \frac{\partial v'}{\partial x_j}, \quad j \in [2, 3] \forall j \in \mathbf{N}. \tag{7}$$

The variation of each term of the $\overline{v'v'}$ budget equation against the $y_I$ was plotted in Fig. 14(a). We notice that all the terms, except the viscous diffusion, are considerable in the $\overline{v'v'}$ budget. The viscous diffusion (*VD*) still remains approximately zero all along the spanwise direction,

$$\mathcal{VD}_{v'v'} = \nu \frac{\partial^2 \overline{v'v'}}{\partial x_j^2}. \tag{8}$$

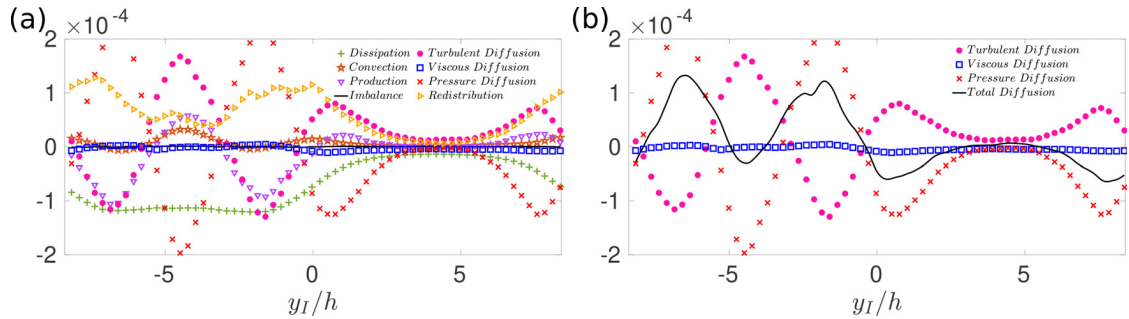

**FIG. 14.** Spanwise variation of (a) different budget terms and (b) various diffusion terms (turbulent diffusion, pressure diffusion, and viscous diffusion) in the budget analysis of Reynolds stress term $\langle \overline{v'v'} \rangle_x$ for the Reynolds number combination (500, 100). The over-line and subscript represent the temporal and spatial x averaging. See Fig. 13 caption for more details.

The convection though non-zero has only minimal contribution. This is due to the fact that $\bar{V}$ and $\bar{W}$ are small nullifying the term's contribution,

$$\mathcal{C}_{v'v'} = -\bar{U}_j \frac{\partial \overline{v'v'}}{\partial x_j}, \quad j = 2, 3. \tag{9}$$

Contrast to the $\overline{u'u'}$ budget, the production given by

$$\mathcal{P}_{v'v'} = -2\left[\overline{v'u'_j} \frac{\partial \bar{V}}{\partial x_j}\right] \tag{10}$$

is no more the significant term in $\overline{v'v'}$ budget since the source of turbulence production here (which is $\frac{\partial \bar{V}}{\partial y}$) is smaller. The most significant contributors in the $\overline{v'v'}$ budget are the diffusion terms—pressure diffusion (PD) and turbulent diffusion (TD) terms,

$$\mathcal{PD}_{v'v'} = -\frac{\partial \overline{v'u'_j v'}}{\partial x_j}; \mathcal{TD}_{v'v'} = -\frac{2}{\rho}\frac{\partial \overline{p'v'}}{\partial y}. \tag{11}$$

Unlike in the $\overline{u'u'}$ budget, the pressure diffusion in the $\overline{v'v'}$ budget does not go to zero since the term is gradient of y here. The two terms though exhibit a similar pattern, they are negatively correlated. However, the overall diffusion (i.e., the sum of pressure, turbulent, and viscous dissipation) is positive which can be seen in Fig. 14(b). The pressure diffusion and turbulent diffusion terms exhibit a sinusoidal variation in the turbulent region. The production being less and diffusion being dominant, the transportation is mainly due to diffusion, and the redistribution process becomes less significant. Further the positive value of redistribution tells that the turbulent energy received is more than the energy distributed. Redistribution or pressure strain-rate is given by

$$\mathcal{PSR}_{v'v'} = 2\left[\frac{p'}{\rho}\frac{\partial v'}{\partial x}\right]. \tag{12}$$

The profile of dissipation in the $\overline{v'v'}$ budget remains qualitatively similar to that of $\overline{u'u'}$ budget; however, the magnitude of dissipation varies according to the magnitudes of production and redistribution. The dissipation of $\overline{v'v'}$ turbulence energy is given as follows:

$$\mathcal{D}_{v'v'} = -2\nu \overline{\frac{\partial v'}{\partial x_j}\frac{\partial v'}{\partial x_j}}. \tag{13}$$

Considering the major players, the budget equation of $\overline{v'v'}$ can be written as

$$\mathcal{D}_{v'v'} + \mathcal{P}_{v'v'} + \mathcal{TD}_{v'v'} + \mathcal{PD}_{v'v'} + \mathcal{PSR}_{v'v'} \approx 0. \tag{14}$$

However, as we enter the non-turbulent region, we see that viscous diffusion, convection, and production drop to near zero at the interface $y_I = 0$ and remain insignificant all along the non-turbulent region (till $y_I = 8.4h$). While the redistribution and dissipation terms gradually decrease, the two diffusion terms show a peculiar trend. The diffusion increases as we enter the non-turbulent region and is maximum at a point close to the interface inside the non-turbulent region. The increase in diffusion terms shows their active role in the transportation of turbulence energy at the interface. Beyond the maximum point, the terms drop quickly to zero and remain insignificant all along the core region. These terms start to grow as they approach the interface $y_I = 8.4h$. With the rise in turbulence energy due to redistribution, the diffusion and dissipation mechanisms start to play their role. The turbulent diffusion and pressure diffusion terms continue to remain anti-correlated in the non-turbulent region. The net negative diffusion [see Fig. 14(b)] close to the interface ($y_I \gtrsim 0$) represents diffusion from neighboring turbulent region to non-turbulent region. The budget equation in the non-turbulent regions truncates to

$$\mathcal{D}_{v'v'} + \mathcal{TD}_{v'v'} + \mathcal{PD}_{v'v'} + \mathcal{PSR}_{v'v'} \approx 0. \tag{15}$$

### 3. $\overline{w'w'}$ budget

To obtain the budget equation for $\overline{w'w'}$, we take $i = k = 3$. The same simplifications are applied which results the equation as follows:

$$0 = -\bar{U}_j \frac{\partial \overline{w'w'}}{\partial x_j} - 2\left[\overline{w'u'_j}\frac{\partial \bar{W}}{\partial x_j}\right] - \frac{\partial \overline{w'u'_j w'}}{\partial x_j} - \frac{2}{\rho}\frac{\partial \overline{p'w'}}{\partial z} + \nu \frac{\partial^2 \overline{w'w'}}{\partial x_j^2}$$
$$+ 2\left[\frac{p'}{\rho}\frac{\partial w'}{\partial z}\right] - 2\nu \overline{\frac{\partial w'}{\partial x_j}\frac{\partial w'}{\partial x_j}}, \quad j \in [2,3] \forall j \in \mathbf{N}. \tag{16}$$

The budget plot of $\overline{w'w'}$ (see Fig. 15) shows that the redistribution is the most dominant term. The predominant flow direction being x and the wave being planar in x and y, and the only source of turbulence in the z direction is redistribution of turbulence from $\overline{u'u'}$ and $\overline{v'v'}$ budget. This makes redistribution the most important term in the budget analysis. The positive value of redistribution tells that the energy is being received. The other terms are quantitatively much lesser. Though the production produces turbulence to some extent, it is much lesser (due to the same reason mentioned in $\overline{v'v'}$ budget case) than the turbulence obtained through redistribution. The terms diffusion and dissipation control the turbulence. The convective part and the viscous diffusion terms are approximately zero. In the non-turbulent region too, all terms drop rapidly to zero. Though there exists turbulence activity close to interface regions ($y_I = 0$ and $y_I = 8.4h$), it can be neglected without affecting the physics.

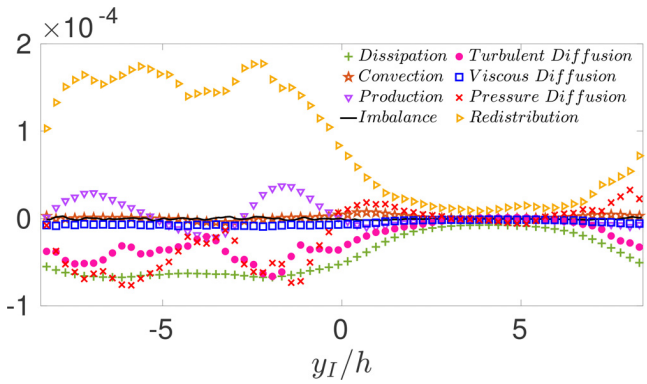

**FIG. 15.** Spanwise variation of different budget terms in the budget analysis of Reynolds stress term $\langle \overline{w'w'} \rangle_x$ for the Reynolds number combination (500, 100). The over-line and subscript represent the temporal and spatial $x$ averaging, respectively. See Fig. 13 caption for more details.

The other components of Reynolds stresses ($\overline{u'v'}, \overline{u'w'}, \overline{v'w'}$) have smaller magnitudes (by an order of magnitude), and, hence, are not presented in the current manuscript for brevity.

### 4. Turbulent kinetic energy

The combination of the three individual Reynolds normal stresses gives turbulent kinetic energy (TKE), which is defined as

$$\bar{k} = \frac{1}{2}(\overline{u'u'} + \overline{v'v'} + \overline{w'w'}). \quad (17)$$

Figure 16 shows the variation of each turbulent quantity along the spanwise direction. We see that the quantities $\langle \overline{v'v'} \rangle_x, \langle \overline{u'u'} \rangle_x$, and hence $\langle \bar{k} \rangle_x$ have significant value in the non-turbulent region. We also notice that the quantities drop and raise gradually in the non-turbulent region forming a "U" shaped variation. As expected, the major contribution to $\langle \bar{k} \rangle_x$ is from $\langle \overline{u'u'} \rangle_x$, followed by $\langle \overline{v'v'} \rangle_x$ and $\langle \overline{w'w'} \rangle_x$. The higher magnitude of $\langle \overline{v'v'} \rangle_x$ compared to $\langle \overline{w'w'} \rangle_x$ could be due to the instability wave.

Figure 16(b) shows the spanwise variation of turbulent kinetic energy ($\langle \bar{k} \rangle_x$) and Reynolds normal stresses ($\langle \overline{u'u'} \rangle_x, \langle \overline{v'v'} \rangle_x, \langle \overline{w'w'} \rangle_x$). A comparison of these quantities between the base state (plotted in blue) and the co-flow state (plotted in red) shows that the TKE is higher in the co-flow state compared to the base state. The enhanced TKE in co-flow is predominantly due to the $\langle \overline{v'v'} \rangle_x$. While the magnitude of $\langle \overline{u'u'} \rangle_x$ is approximately the same in both the cases, the magnitude of $\langle \overline{w'w'} \rangle_x, \langle \overline{v'v'} \rangle_x$ is higher in co-flow compared to base flow. The quantities are almost twice their base state magnitude on the turbulent side and half the base state magnitude on the non-turbulent side. We believe that the enhancement of $\langle \overline{v'v'} \rangle_x$ in the co-flow state is due to the shear layer instability. However, the increase in $\langle \overline{w'w'} \rangle_x$ may not be due to the instability wave since the wave is planar (existing in the $xy$-plane). The increase in $\langle \overline{w'w'} \rangle_x$ is possibly due to the additional redistribution in co-flow.

Figure 17(a) shows the spanwise variation of $\langle \bar{k} \rangle_x$ at different wall-normal locations. We observe the monotonic increase in $\langle \bar{k} \rangle_x$ on the turbulent side with $z$ till $z = 1.55h$, which drops beyond $z = 1.55h$. The TKE at $z = 1.79h$ is less than that at $z = 1.55h$. As we move along $z$, the drop between turbulent and non-turbulent regions becomes steeper and close to the top wall, and TKE drops abruptly at the interface. This could be due to the decrease in the width of the mixing layer with $z$ (see Fig. 5), which is elucidated in Sec. III A 1.

Furthermore, the article has so far discusses the statistics for a specific Reynolds number ratio [$r = 5$ corresponding to Reynolds number combination (500,100)]. However, previous study by Teja et al.[20] has shown that the Reynolds number ratio has considerable effect on flow physics. To explore the effect of "$r$" on turbulent kinetic energy, Fig. 17(b) was plotted to understand the spanwise variation of $x-$ averaged turbulent kinetic energy ($\langle \bar{k} \rangle_x$) for different Reynolds number ratios ($r = Re_h/Re_l$) 10, 7.5, 5, 3.5, and 2, corresponding to Reynolds number combinations ($Re_h$, $Re_l$) (500,50), (500,66), (500,100), (500,143), and (500,250). We see that cases $r = 10$ and $r = 7.5$ have the same magnitude, and the cases $r = 3.5$ and $r = 2$ have an approximately same magnitude. The kinetic energy of $r = 5$ case has an intermediate value.

To investigate the kinetic energy deeper, a budget analysis of the turbulent kinetic energy ($\bar{k}$) is done. Figure 18 shows the spanwise variation of various budget terms of $\bar{k}$. We see that the turbulent diffusion has no contribution to the diffusion process at the interface regions but plays a significant role at the center of the turbulent flow region far away from the turbulent/non-turbulent interface regions. Close to the interface, pressure diffusion has a significant role. At the points of interface $y_I = 8.4h$ or $-8.4h$, and $y_I = 0$, we see that the pressure diffusion takes a negative value, which means that the process tends to diffuse turbulence from the turbulent region to the non-turbulent region. The pressure diffusion exhibits a wavy variation in the turbulent region. The positive value of pressure diffusion tells the transportation of turbulence away from the interface region. Given the negligible production of $\overline{v'v'}$ and $\overline{w'w'}$, the production of kinetic energy is primarily due to the $\overline{u'u'_j} \frac{\partial \bar{U}}{\partial x_j}$ ($P_{\overline{u'u'}}$). Like in every budget analysis, viscous dissipation has no significant role in the budget analysis of turbulent kinetic energy.

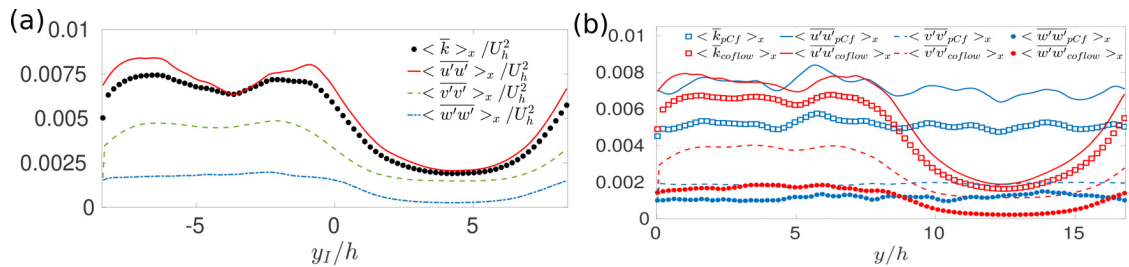

**FIG. 16.** (a) Spanwise variation of turbulent kinetic energy ($\langle \bar{k} \rangle_x$) and the Reynolds stress terms ($\langle \overline{u'u'} \rangle_x, \langle \overline{v'v'} \rangle_x, \langle \overline{w'w'} \rangle_x$) in a plane Couette flow (pCf); (b) comparison of quantities plotted in (a) between pCf and co-flow. The plots are plotted at mid $z$-location ($z = 1h$). The quantities in both the figures are time averaged and $x$ averaged and are non-dimensionalized with $U_h^2$. See Fig. 13 caption for more details.

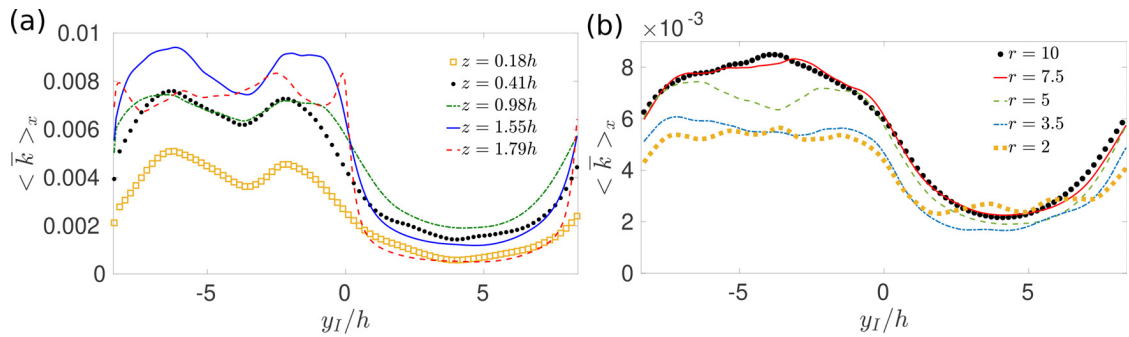

**FIG. 17.** Spanwise variation of turbulent kinetic energy ($\langle \bar{k} \rangle_x$) at different (a) wall-normal locations (z) and (b) Reynolds number ratios (r). The overline and subscript represent the temporal and spatial averaging, respectively.

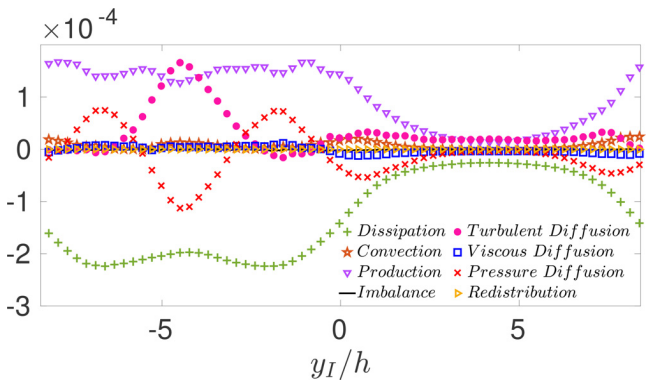

**FIG. 18.** Spanwise variation of different budget terms in the budget analysis of turbulent kinetic energy ($\langle \bar{k} \rangle_x$) for the Reynolds number combination (500, 100). The over-line and subscript represent the temporal and spatial x averaging, respectively. See Fig. 13 caption for more details.

## IV. CONCLUSION

This article presents a detailed analysis on co-flowing turbulent/non-turbulent plane Couette flows and the effect of one flow on the other. The enstrophy jump observed at the interface confirms that the jump is a characteristic feature of TNTI, irrespective of class of flow. The enstrophy contours at different wall-normal planes confirm the transportation of vorticity along the wall-normal direction. Comparing the vortical structures between the co-flow and standard pCf, the vortices are broader and longer in the co-flow case, signifying the excessive straining process. This was also confirmed through vortex straining iso-contours. The geometric and topological features were analyzed using invariant maps. As expected, the channel core predominantly has vortical structures, and the dissipative structures appear close to the walls. However, it was found that the vortices have sheet-like structures irrespective of wall-normal distance in both standard pCf and co-flow.

The analysis of first and second-order quantities showed that the non-turbulent flow exhibits higher mean velocity than what one expects analytically. This additional momentum further varies along the spanwise location. The momentum is higher close to the interface and decreases as one moves into the non-turbulent region away from the interface. The effect of non-turbulent flow on the turbulent flow can also be seen from the $\overline{u'u'}$ profile. While the profile is symmetric about mid $z-$ plane for a conventional pCf, it becomes skewed toward the top plate in the co-flowing pCf scenario. Furthermore, the non-turbulent region has a significant Reynolds stress ($\overline{u'u'}$), indicating the transportation of turbulence from turbulent flow to non-turbulent flow. To dwell into the transportation mechanism of turbulence, a budget analysis of $\overline{u'u'}$ is done, which leads to some interesting observations. The redistribution and production process played a dominant role. Similarly, budget analysis carried out for $\overline{v'v'}$ showed that the diffusion has a dominant role in transporting turbulence to the non-turbulent region. The budget analysis of $\overline{w'w'}$ shows that the only source of turbulence in the wall-normal direction is redistribution. The turbulent kinetic energy (k) budget analysis revealed the local dominance of pressure and turbulent diffusion. However, most of the total energy contained in the flow is still contributed by convection.

## ACKNOWLEDGMENTS

This work has received support from the P.G. Senapathy Center for Computing Resource, IIT Madras, through a grant of computing time.

## AUTHOR DECLARATIONS
### Conflict of Interest

The authors have no conflicts to disclose.

**Author Contributions**

**Manohar Teja Kalluri:** Data curation (equal); Formal analysis (equal); Investigation (equal); Methodology (equal); Validation (equal); Visualization (equal); Writing – original draft (equal). **Vagesh D. Narasimhamurthy:** Conceptualization (equal); Funding acquisition (equal); Methodology (equal); Project administration (equal); Resources (equal); Supervision (equal); Writing – review & editing (equal).

## DATA AVAILABILITY

The data that support the findings of this study are available from the corresponding author upon reasonable request.

## APPENDIX A: VORTICITY EQUATION

The vorticity equation is obtained taking the curl of Navier–Stokes equation. A simplified form of vorticity equation is given as follows:

$$\frac{\partial \omega_i}{\partial t} + U_j \frac{\partial \omega_i}{\partial x_j} = -\omega_i \frac{\partial U_j}{\partial x_j} + \epsilon_{ijk} \frac{1}{\rho^2} \frac{\partial \rho}{\partial x_j} \frac{\partial p}{\partial x_k} + \omega_j \frac{\partial U_i}{\partial x_j} + \nu \frac{\partial^2 \omega_i}{\partial x_j^2}. \quad \text{(A1)}$$

The first term on the left-hand side (LHS) represents the temporal variation of vorticity. The second term is called the convective term, which represents the transport of strained vortex. The first term on the right-hand side (RHS) represents the expansion of the vorticity field. This term is present only for a compressible flow scenario and becomes zero for an incompressible flow case because of continuity condition. The second term on the RHS is the torque/baroclinic term. This term is responsible for the generation of vorticity. However, this term exists only for stratified flow conditions. The current flow system has a constant density, making the spatial gradient of density zero, thus making the baroclinic torque term zero. The third term is the vortex stretching term. This shows the physical mechanism of elongation of a vortex due to the spatial gradient of velocity. The last term on the RHS shows the viscous diffusion of vorticity. The vorticity equation for a baroclinic, incompressible flow such as the current case simplifies to

$$\frac{\partial \omega_i}{\partial t} + U_j \frac{\partial \omega_i}{\partial x_j} = \omega_j \frac{\partial U_i}{\partial x_j} + \nu \frac{\partial^2 \omega_i}{\partial x_j^2}. \quad \text{(A2)}$$

## APPENDIX B: THEORY OF FLOW INVARIANTS

The invariants are robust tools for the analysis since they are intrinsic parts of the flow system and independent of the coordinate-axis. We begin with shedding light on the procedure to obtain the invariants and later use them to obtain information about the flow structures, and their deformations by plotting the invariant maps. Here on, we use the notation $\phi_{i,j}$ to indicate the partial differentiation in tensoral form, i.e., $\phi_{i,j} = \frac{\partial \phi_i}{\partial x_j}$, where $\phi$ is any variable. The deformation matrix ($U_{i,j}$) can be split into two terms—a symmetric term (rate of strain tensor, $S_{ij}$) and an anti-symmetric term (rate of rotation tensor, $\Omega_{ij}$).

$$U_{i,j} = S_{ij} + \Omega_{ij}, \quad \text{(B1a)}$$

$$S_{ij} = \frac{1}{2}(U_{i,j} + U_{j,i}), \quad \text{(B1b)}$$

$$\Omega_{ij} = \frac{1}{2}(U_{i,j} - U_{j,i}). \quad \text{(B1c)}$$

### 1. Obtaining the invariants

The invariants for a second order tensor $M_{3\times3}$ are calculated by taking the determinant of characteristic matrix ($|M - \lambda I|$, where $I$ is the identity matrix of order $3 \times 3$ and $\lambda$ is a variable). Equating the determinant of characteristic matrix to zero results in a cubic equation in terms of $\lambda$, whose solution gives the eigenvalues of the matrix $M$.

### 2. Invariants of velocity gradient tensor

The velocity gradient tensor $U_{i,j}$ is a second order tensor ($U_{3\times3}$). Invariants for the same are obtained as explained before (see Appendix B). Alternately, the invariants can be calculated directly using the following formulas:[11]

$$P = -\text{trace}(U_{i,j}) = -U_{i,i}, \quad \text{(B2a)}$$

$$Q = -\frac{1}{2} U_{i,j} U_{j,i} = \frac{1}{4}(\omega_i \omega_i - 2 S_{ij} S_{ij}), \quad \text{(B2b)}$$

$$R = -\frac{1}{3} U_{i,j} U_{j,k} U_{k,i} = -\frac{1}{3}\left(S_{ij} S_{jk} S_{ki} + \frac{3}{4} \omega_i \omega_j S_{ij}\right), \quad \text{(B2c)}$$

where $\omega_i$ and $\omega_j$ represent the $i$th and $j$th vorticity components, respectively. The topological features of the velocity gradient tensor in the (P,Q,R) space have been explained in detail by Chong et al.[40] The surface given by the equation $27R^2 + (4P^3 - 18PQ)R + (4Q^3 - P^2Q^2) = 0$ divides the (P,Q,R) space into two regions: one region having two complex and one real eigenvalues and other having three real eigenvalues. However, when the flow is incompressible, the first invariant $P$ becomes zero because of continuity, and the three dimensional space becomes two dimensional; the data Q and R on the plane $P = 0$ are studied. This simplifies the three dimensional surface equation to a two dimensional equation $27R^2 + 4Q^3 = 0$. The value $27R^2 + 4Q^3$ is called the discriminant ($D$). This discriminant line $D = 0$ splits the (Q,R) plane into two regions. Any point in the region above the discriminant line has two complex and one real eigenvalue, while the region below the discriminant line has three distinct real eigenvalues. A point on the discriminant line has three real eigenvalues of which two are equal. Figure 19(a) shows the pictorial description of the same. A further classification is done based on the sign of R. The points in the region $R < 0$ have complex eigenvalues with negative real part and are termed as stable. The points in the region $R > 0$ have complex eigenvalues with the positive real part and are termed as unstable. The points above discriminant line are termed focus, and points below discriminant line are called saddle. The classification based on the sign of R is explained in Fig. 19(b). An elaborated explanation on the physical meaning of terms stable, unstable, focus, and saddle can be read from Chong et al.[40]

The discriminant line and $R = 0$ line divide the (Q,R) plane into four quadrants: $R > 0, D > 0 (Q_1); R < 0, D > 0 (Q_2); R < 0, D < 0 (Q_3)$; and $R > 0, D < 0 (Q_4)$. Each quadrant tells us the deformation that the fluid element is undergoing and the structure of the fluid element. Figure 20 shows the pictorial representation of the same.

The mathematical expressions of the quantities Q and R are examined keenly to understand their physical interpretation. The strain rate product $S_{ij}S_{ij}$ represents the viscous dissipation process ($\varepsilon = 2\nu S_{ij}S_{ij}$). Thus, the second invariant Q compares the viscous dissipation and enstrophy. A negative Q value represents the local domination of strain product over enstrophy, while a positive value of $Q$ represents the local domination of enstrophy over strain product. Hence, the regions with positive $Q$ value provide us information about vortex structures, and the regions with negative $Q$ value provide us information about dissipative structures. The sign of $R$ depends on $S_{ij}$. A positive $S_{ij}$ results in a negative $R$ value, which signifies elongation (positive strain). Similarly, negative $S_{ij}$ makes the sign of $R$ positive, which signifies compression. Thus, a positive $Q$ and negative $R$ ($Q_2$) represent the vortex stretching, and a positive $Q$ and positive $R$ ($Q_1$) represent the vortex compression. On the other hand, when $Q$ is less than zero, the region is dominated by viscous dissipation. The structures of intense values of viscous dissipation tend to appear in the form of sheets and ribbons in an isotropic turbulence.[41] The regions with positive $R$ have sheet-like structures, and the regions with negative $R$ have tube-like structures. Thus, the ($Q$, $R$) invariant map tells us the relation between the flow topology and the dissipation/production of enstrophy by vortex stretching/compression.

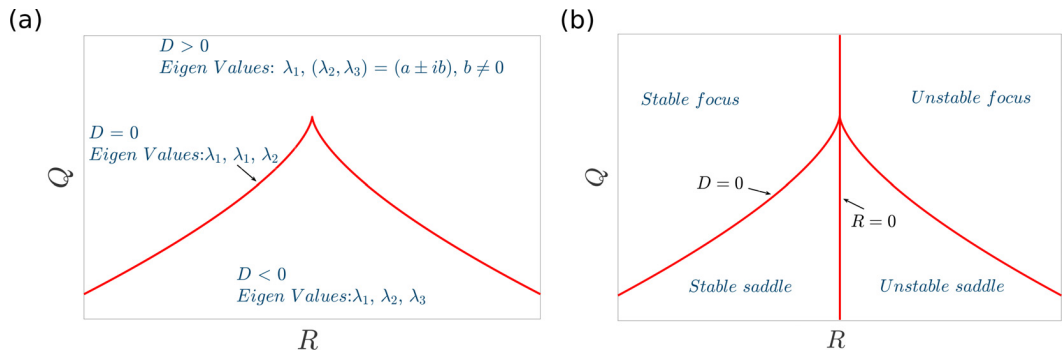

**FIG. 19.** (a) Nature of eigenvalues based on the sign of discriminant ($D = 27R^2 + 4Q^3$) and (b) nature of point depending on the ($Q$, $R$) coordinate location.

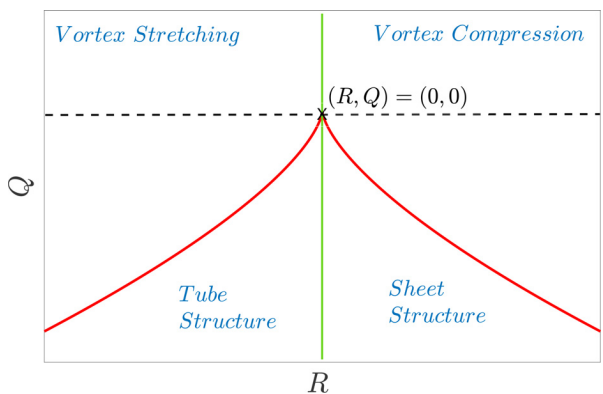

**FIG. 20.** Topological flow features shown by flow structures based on the ($R$, $Q$) values.

### 3. Invariants of rate of strain tensor

The invariants for the rate of strain tensor [see Eq. (B1b)] are obtained as explained in Appendix B 1. Alternately, the invariants can be obtained by replacing $U_{i,j}$ with $S_{ij}$, since $\Omega_{ij} = 0$ for a strain flow. Thus, the invariants for $S_{ij}$ become

$$P_s = \text{trace}(S_{ij}) = S_{ii}, \quad \text{(B3a)}$$

$$Q_s = -\frac{1}{2}S_{ij}S_{ij}, \quad \text{(B3b)}$$

$$R_s = -\frac{1}{3}(S_{ij}S_{jk}S_{ki}). \quad \text{(B3c)}$$

The first invariant of rate of strain tensor becomes zero in the case of incompressible fluid because of continuity. The second invariant is related to dissipation of instantaneous kinetic energy ($\varepsilon = -2\nu S_{ij}S_{ij} = 4\nu Q_s$). Thus, the regions of high $Q_s$ have high dissipation. The third invariant represents the skewness of strain. The term corresponds to the production term of the transport equation of $S_{ij}S_{ij}$,[11] which is given by

$$\frac{D}{Dt}\left(\frac{1}{2}S_{ij}S_{ij}\right) = -S_{ij}S_{jk}S_{ki} - \frac{1}{4}\omega_i\omega_j S_{ij} - S_{ij}\frac{\partial^2 p}{\partial x_i \partial x_j} + \nu S_{ij}\nabla^2 S_{ij}. \quad \text{(B4)}$$

Thus, a positive value of $R_s$ corresponds to production of strain product term ($S_{ij}S_{ij}$), and a negative value of $R_s$ corresponds to destruction of strain product. Higher values of $R_s$ are also the regions of more viscous dissipation and, hence, imply the presence of structures of form sheets and ribbons as mentioned previously. The positive value of $R_s$ corresponds to sheet-like structures, and negative $R_s$ corresponds to tube-like structures. While the structures in the region with positive $R_s$ undergo expansion, the structures in the region with negative $R_s$ compress.[11] Figure 21(a) shows the above information pictorially. Thus, the invariant map of ($R_s$, $Q_s$) explains the geometry of straining of fluid elements.

### 4. Invariants of rate of rotation tensor

The invariants for the rate of rotation tensor [see Eq. (B1c)] are obtained as explained in Appendix B 1. Alternately, the invariants can be obtained by replacing $U_{i,j}$ with $S_{ij} + \Omega_{ij}$ and taking $S_{ij} = 0$. Thus, the invariants for $\Omega_{ij}$ become

$$P_\Omega = \text{trace}(\Omega_{ij}) = 0, \quad \text{(B5a)}$$

$$Q_\Omega = -\frac{1}{2}\omega_i\omega_i, \quad \text{(B5b)}$$

$$R_\Omega = 0, \quad \text{(B5c)}$$

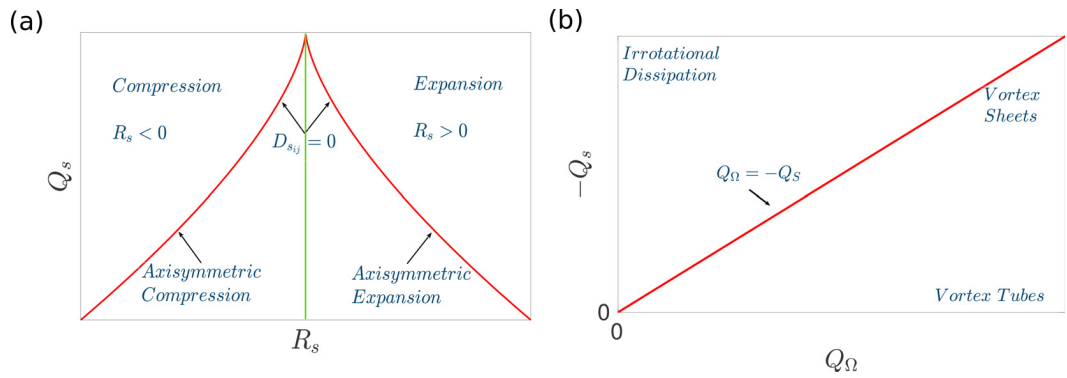

**FIG. 21.** (a) Geometric deformations undergone by fluid elements based on the ($R_s$, $Q_s$) combination and (b) nature of flow structures based on ($Q_\Omega$, $Q_s$) values.

where $P_\Omega$ become zero as the principal diagonal elements ($\Omega_{ii}$) are zeroes. The second invariant is closely related to enstrophy ($\omega_i \omega_i$). Thus, enstrophy can also be written as $-2Q_\Omega$, and, hence, the regions of high/low $Q_\Omega$ have high/low enstrophy values, respectively. The invariant map of ($Q_w$, $-Q_s$) can be used to explain the topology of dissipation of kinetic energy. The points aligned along the horizontal line represent large values of $Q_\Omega$ and represent regions of high enstrophy and negligible dissipation, and the points aligned along the vertical line represent large values of $Q_s$ and represent strong dissipation ("irrotational dissipation"). The points with equally high magnitudes of $Q_w$ and $Q_s$ are observed in the vortex sheet structures.[11]

## APPENDIX C: BUDGET EQUATION OF A REYNOLDS STRESS ($\overline{u'_i u'_k}$)

The budget equation for any given stress component ($\overline{u'_i u'_k}$) can be obtained as

$$\frac{\partial \overline{u'_i u'_k}}{\partial t} = -\bar{U}_j \frac{\partial (\overline{u'_i u'_k})}{\partial x_j} - \left[\overline{u'_i u'_j}\frac{\partial \bar{U}_k}{\partial x_j} + \overline{u'_k u'_j}\frac{\partial \bar{U}_i}{\partial x_j}\right] - \frac{\partial \overline{u'_i u'_j u'_k}}{\partial x_j}$$
$$- \frac{1}{\rho}\frac{\partial (\overline{p'u'_i}\delta_{kj} + \overline{p'u'_k}\delta_{ij})}{\partial x_j} + \nu \frac{\partial^2 \overline{u'_i u'_k}}{\partial x_j^2} + \frac{\overline{p'}}{\rho}\left(\frac{\partial u'_k}{\partial x_i} + \frac{\partial u'_i}{\partial x_k}\right)$$
$$- 2\nu \overline{\frac{\partial u'_i}{\partial x_j}\frac{\partial u'_k}{\partial x_j}}. \tag{C1}$$

The term on the LHS of Eq. (C1) is the *temporal term*, which remains only for an unsteady case and disappears when the flow is steady. The first term on the right-hand side (RHS) is the *convection term*, which signifies the spatial variation of $\overline{u'_i u'_k}$ for a moving fluid element. The second and third RHS terms (given together in the square brackets) represent the interaction between the turbulence quantities and the mean flow quantities. The gradient of mean flow velocity acts on the turbulence terms. This process helps to sustain the turbulence, which otherwise slowly dies. Hence, the term is called the *production term*. The fourth term on the RHS of the equation, (which could also be written as $-u'_j \frac{\partial (u'_i u'_k)}{\partial x_j}$), represent, the diffusion of $\overline{u'_i u'_k}$ along the $j$th direction due to fluctuating velocity $u'_j$. Hence, the term is called the *turbulent diffusion*. The fifth term on the RHS represents the diffusion of turbulence due to fluctuating pressure and is called the *pressure diffusion*. The sixth term that could be written as $\nu \frac{\partial}{\partial x_j}\left(\frac{\partial \overline{u'_i u'_k}}{\partial x_j}\right)$ represents the diffusion due to viscosity, and, hence, the term is called the *viscous diffusion*. The seventh RHS term shows the interaction between fluctuating pressure and the fluctuating strain rate. This term is called the *pressure strain rate* (PSR) or the *redistribution* term. This term distributes turbulent energy from one direction to another. The last term is called the *dissipation* term, which represents the dissipation of turbulent energy due to viscosity. Substituting natural numbers between 1 and 3 for the free indices $i$ and $k$ in the above Eq. (C1), we can get conservation equation for various stress terms $\overline{u'_i u'_k}$.